\documentstyle[12pt,bezier]{article}
\begin{document}
\newcommand{\uuu}{\mbox{\huge $u$}}
\newcommand{\vvv}{\mbox{\huge $v$}}
\newcommand{\xxx}{\end{document}}
\newcommand{\cue}{{\cal E}}
\newcommand{\cun}{{\mbox{\scriptsize${\cal N}$}}}
\newcommand{\bcc}{\begin{center}}
\newcommand{\ecc}{\end{center}}
\def \inbar{\vrule height1.5ex width.4pt depth0pt}
\def \C{\relax\hbox{\kern.25em$\inbar\kern-.3em{\rm C}$}}
\def \R{\relax{\rm I\kern-.18em R}}
\newcommand{\Z}{\ Z \hspace{-.08in}Z}
\newcommand{\be}{\begin{equation}}
\newcommand{\ee}{\end{equation}}
\newcommand{\bea}{\begin{eqnarray}}
\newcommand{\eea}{\end{eqnarray}}
\newcommand{\p}{\psi}
\newcommand{\f}{\phi}
\newcommand{\g}{\gamma}
\newcommand{\G}{\Gamma}
\newcommand{\e}{\eta}
\newcommand{\m}{\mu}
\newcommand{\n}{\nu}
\newcommand{\s}{\sigma}
\renewcommand{\t}{\tau}
\renewcommand{\d}{\delta}
\renewcommand{\pt}{\frac{\partial}{\partial t}}
\newcommand{\ppt}{\frac{\partial^{2}}{\partial t^{2}}}
\newcommand{\nn}{\nonumber}
\newcommand{\kt}{\rangle}
\newcommand{\br}{\langle}
\newcommand{\fs}{\small}
\newcommand{\so}{S_{0}}
\newcommand{\I}{\mbox{1}_{m\times m}}
\newcommand{\In}{\mbox{1}_{n\times n}}
\newcommand{\xo}{x_{0}}
\newcommand{\po}{\psi_{0}}
\newcommand{\eo}{\e_{0}}
\newcommand{\ts}{\tilde{S}}
\newcommand{\pss}{\frac{\partial}{\partial s}}
\newcommand{\tcuf}{\tilde{\cal F}}
\newcommand{\cuf}{{\cal F}}
\newcommand{\oot}{\mbox{\fs$\frac{1}{2}$}}
\newcommand{\iot}{\mbox{\fs$\frac{i}{2}$}}
\newcommand{\cur}{{\cal R}}
\newcommand{\iv}{\imath\! v}
\newcommand{\ib}{\int_{0}^{\b}}
\newcommand{\lll}{\left( }
\newcommand{\rrr}{\right)}
\newcommand{\llc}{\left\{ }
\newcommand{\rrc}{\right\} }
\newcommand{\lpt}{\left.}
\newcommand{\rpt}{\right.}
\newcommand{\rar}{\longrightarrow}
\newcommand{\lar}{\longleftarrow}
\newcommand{\cinf}{C^{\infty}\!}
\newcommand{\vol}{d\mbox{\bf\fs$\Omega$}}
\newcommand{\sx}{\mbox{\fs$(x)$}}
\newcommand{\hD}{\hat{D}}
\newcommand{\V}{(V)}
\newcommand{\La}{\Lambda}
\newcommand{\ph}{{\cal P}({\cal H})}
\newcommand{\cp}{\C\! P}
\newcommand{\Psusy}{Parasupersymmetric}
\newcommand{\psusy}{parasupersymmetric}
\newcommand{\tii}{topological invariant}
\newcommand{\tiis}{topological invariants}
\newcommand{\Tii}{Topological invariant}
\newcommand{\qq}{{\cal Q}}
\newcommand{\ddd}{\Delta^{(p=2)}}
\newcommand{\psqm}{($p=2$)--PSQM}
\title{Topological Aspects of Parasupersymmetry}
\author{Ali Mostafazadeh\\ \\ Department of Physics,
Sharif University of Technology\\ P.~O.~Box 11365-9161, Tehran, Iran, and\\
Institute for Studies in Theoretical Physics and Mathematics\\
P.~O.~Box 19395-1795, Tehran, Iran.}
\maketitle
\abstract{
\Psusy~ quantum mechanics is exploited to introduce a \tii~ associated with
a pair of parameter dependent Fredholm (respectively elliptic
differential) operators satisfying two compatibility
conditions. An explicit algebraic expression for this \tii~ is provided.
The latter identifies the parasupersymmetric topological invariant with the
sum of the analytic (Atiyah-Singer) indices of the corresponding operators.}

\section{Introduction}
Perhaps one of the most intriguing aspects of supersymmetry is its
relation with the Atiyah-Singer index theorem \cite{as}. It was Witten
\cite{w1} who first recognized this relation in the context of
supersymmetric quantum mechanics (SQM). The subsequent developments
in this direction have led to supersymmetric proofs of the index theorem
\cite{a-g-w}.

During the same period of development of supersymmetric
proofs of the index theorem, i.e., mid 1980's, Rubakov and Spiridonov
(R-S) \cite{rs}
introduced their ($p=2$)-parasupersymmetric quantum mechanics (PSQM). This
involved a generalization of the superalgebra of SQM (see
Eq.~(\ref{q1}) below), namely the {\em parasuperalgebra}:
        \bea
        \qq^3\:=\:0\;,~\; [\qq ,H]&=& 0 \label{q0.1}\;, \\
        \{\qq^2,\qq^\dagger\}+\qq\qq^\dagger\qq&=&4\qq H\;.
        \label{q0.2}
        \eea
The defining parasuperalgebra (\ref{q0.1}), (\ref{q0.2}) have since
been generalized to arbitrary order $p>2$, by Khare \cite{kh1}, and modified
by Beckers and Debergh (B-D) \cite{bd}. B-D ($p=2$)-parasuperalgebra is
given by Eqs.~(\ref{q0.1}) and
        \be
        \left[ \qq ,\left[\qq^\dagger, \qq\right]\right]=2\qq H\;.
        \label{q0.4}
        \ee

In a preceding article \cite{ali}, it is shown that a careful analysis of
the defining parasuperalgebra (for both R-S and B-D types) provides
important information on the degeneracy structure of the spectrum of
the corresponding systems. In particular, postulating the existence of
a parasupersymmetry involution (chirality) operator and supplementing
(either of) the parasuperalgebra(s) with an additional relation
expressing the Hamiltonian in terms of the parasupercharges, namely
        \be
        H=\frac{1}{2}\left[
        (\qq\qq^\dagger)^2+(\qq^\dagger\qq)^2-
        \frac{1}{2}(\qq\qq^{\dagger 2}\qq+\qq^\dagger \qq^2
        \qq^\dagger)\right]^\frac{1}{2}\;,
        \ee
one can show that the integer
        \bea
        \Delta^{(p=2)}&:=&n^{\pi B} - 2n^{\pi F}\: =\:
        n^{\pi B}_0 -2n^{\pi F}_0\;, \label{q0.5}\\
        n^{\pi B}&:=&\mbox{number of parabosonic states}\nn\\
        n^{\pi F}&:=&\mbox{number of parafermionic states}\nn \\
        n^{\pi B}_0&:=&\mbox{number of zero energy parabosonic states}\nn\\
        n^{\pi F}_0&:=&\mbox{number of zero energy parafermionic states}\nn
        \eea
is a topological invariant. Furthermore, it is shown in \cite{ali}
that $\Delta^{(p=2)}$ is a measure of parasupersymmetry breaking, i.e., the
condition $\Delta^{(p=2)}\neq 0$ implies the exactness of parasupersymmetry.
In this respect, it is quite similar to the Witten index of supersymmetry.

The purpose of the present letter is to explore the mathematical meaning
of $\Delta^{(p=2)}$. In section 2, a brief discussion of SQM is
presented to demonstrate the motivation for the proceeding analysis of PSQM.
In section 3, a derivation of the expression for $\Delta^{(p=2)}$
is offered and the main result of the letter is presented.
Section~4 includes the concluding remarks.

\section{SQM and the Index Theorem}
The main ingredient of SQM which makes its relation with the index theory
possible, is its simple degeneracy structure. More precisely, the
degeneracy structure of the spectrum of any supersymmetric quantum
mechanical system is determined using only the defining superalgebra:
        \be
        \qq^2=0,\; [\qq,H]\:=\:0,\; \{\qq,\qq^\dagger\}\:=\:
        2H\,,\label{q1}
        \ee
and the properties of the supersymmetry involution (chirality) operator
$\tau$:
        \be
        \tau^2=1,\; \tau^\dagger=\tau,\; \{\qq,\tau\}=0\;.
        \label{q2}
        \ee
In Eqs.~(\ref{q1}) and~(\ref{q2}), $\qq$ stands for (one of) the
generator(s) of supersymmetry, $\qq^\dagger$ is its adjoint, and
$H$ is the Hamiltonian. The chirality operator $\tau$ induces a
double grading of the Hilbert space, ${\cal H}={\cal H}_+\oplus{\cal H}_-$,
where
        \be
        {\cal H}_\pm :=\left\{ \psi\in{\cal H}\: :\: \tau\psi=\pm\psi
        \right\}\;.\label{q3}
        \ee
The superalgebra (\ref{q1}) can be employed to show that the energy
spectrum is non-negative and that each positive energy state of
definite chirality is accompanied with another state of the same
energy and opposite chirality, \cite{w1,ali}. In this sense, one says
that the positive energy levels are {\em doubly degenerate}.

Introducing the self-adjoint generators:
        \be
        Q_1=\frac{1}{\sqrt{2}}(\qq+\qq^\dagger)\;,~~~~
        Q_2=\frac{-i}{\sqrt{2}}(\qq-\qq^\dagger)\;,
        \label{q4}
        \ee
one rewrites the superalgebra (\ref{q1}) in the form:
        \bea
        \{ Q_1,Q_2\}&=& 0\;,\label{q1.1}\\
        Q_1^2\:=\: Q_2^2&=&H\;,\label{q1.2}\\
        \left[ Q_1,H\right]&=&0\;,\label{q1.3}\\
        \left[ Q_2,H\right]&=&0\;,\label{q1.4}\\
        \{Q_1,\tau\}&=&0\;,\label{q1.5}\\
        \{Q_2,\tau\}&=&0\;,\label{q1.6}\\
        \tau^2\:=\:1&,&\;\tau^\dagger\:=\:\tau\;.\label{q1.7}
        \eea
In view of (\ref{q1.3}), one can use the eigenvalues $E$ and
$q_1=\pm\sqrt{E}$ of $H$ and $Q_1$, to label the states. Here we
choose not to include any other quantum numbers. Their presence
will not interfere with the arguments presented in this letter.

For each positive energy level $(E>0)$,the $\{ |E,\pm\sqrt{E}\kt\}$ basis
may be used to yield matrix representations of the relevant operators
\cite{ali}. Let us denote by ${\cal H}_E$ the eigenspace associated with
the eigenvalue $E$, then
        \bea
        \left. Q_1\right|_{{\cal H}_E}=\sqrt{E}\left( \begin{array}{cc}
        1&0\\
        0&-1\end{array}\right)=\sqrt{E}\sigma_3\;,&&
        \left. Q_2\right|_{{\cal H}_E}=\sqrt{E}\left( \begin{array}{cc}
        0&1\\
        1&0\end{array}\right)=\sqrt{E}\sigma_1,\nn\\
        \left. \tau\right|_{{\cal H}_E}=\left( \begin{array}{cc}
        0&-i\\
        i&0\end{array}\right)=\sigma_2,&&
        \left. H\right|_{{\cal H}_E}=E\left( \begin{array}{cc}
        1&0\\
        0&1\end{array}\right)\;,\nn
        \eea
where $\sigma_i$, $i=1,2,3$, are Pauli matrices.
Let us transform into a basis where $\tau$ and $H$ are diagonal. In such
a basis:
        \bea
        \left. Q_1\right|_{{\cal H}_E}&=&\sqrt{E}\left( \begin{array}{cc}
        0&-i\\
        i&0\end{array}\right)=\sqrt{E}\sigma_2\;,~~~~~~~~~~~~~~~~\nn\\
        \left. Q_2\right|_{{\cal H}_E}&=&\sqrt{E}\left( \begin{array}{cc}
        0&1\\
        1&0\end{array}\right)=\sqrt{E}\sigma_1,\label{q5}~~~~~~~~~~~~~~~~\\
        \left. \tau\right|_{{\cal H}_E}=\left( \begin{array}{cc}
        1&0\\
        0&-1\end{array}\right)&=&\sigma_3,~~~~~~~
        \left. H\right|_{{\cal H}_E}=E\left( \begin{array}{cc}
        1&0\\
        0&1\end{array}\right)\;.\nn
        \eea
The fact that $trace(\left. \tau\right|_{{\cal H}_E})=0$ is the very
reason for the topological invariance of the Witten index \cite{w1}:
        \bea
        {\rm index}_W&:=&{\rm trace}(\tau)\:=\:
        n^B-n^F\:=\:n^B_0-n^F_0\;, \label{q6}\\
        n^{B}&:=&\mbox{number of bosonic states}\nn\\
        n^{F}&:=&\mbox{number of fermionic states}\nn \\
        n^{B}_0&:=&\mbox{number of zero energy bosonic states}\nn\\
        n^{F}_0&:=&\mbox{number of zero energy fermionic states}\nn
        \eea

Eq.~(\ref{q5}) serves as a motivation for relating the Witten index with
the analytic indices of Fredholm operators. To demonstrate this relationship,
first one introduces the representation
        \be
        {\cal H}=\left(\begin{array}{c}
        {\cal H}_+\\
        {\cal H}_-\end{array}\right)
        \;\label{q7}
        \ee
of the Hilbert space in which $\tau$ is (block-)diagonal. To obtain
the representations of $Q_i$ ($i=1,2$), one appeals to Eqs.~(\ref{q1.5})
and (\ref{q1.6}). These together with (\ref{q5}) suggest:
        \be
        Q_1=\left(\begin{array}{cc}
        0&-iD_1^\dagger\\
        iD_1&0\end{array}\right)\;\;,
        Q_2=\left(\begin{array}{cc}
        0&D_2^\dagger\\
        D_2&0\end{array}\right)\;,
        \label{q8}
        \ee
where $D_i:{\cal H}_+\to{\cal H}_-$, $i=1,2$ are a couple of operators
acting on ${\cal H}_+$ and $D_i^\dagger$ are their adjoints. Enforcing
the superalgebra, namely Eqs.~(\ref{q1.1}) and (\ref{q1.2}), this
representation  leads to the following set of compatibility conditions
for $D_i$:
        \bea
        D_1^\dagger D_2&=&D_2^\dagger D_1 \;,\label{q9}\\
        D_1 D_2^\dagger &=&D_2 D_1^\dagger \;,\label{q10}\\
        D_1^\dagger D_1&=&D_2^\dagger D_2 \;,\label{q11}\\
        D_1 D_1^\dagger &=&D_2 D_2^\dagger \;.\label{q12}
        \eea
In view of Eqs.~(\ref{q1.2}), (\ref{q11}), and (\ref{q12}), the
Hamiltonian takes the form:
        \be
        H=\left(\begin{array}{cc}
        D_i^\dagger D_i&0\\
        0&D_i D_i^\dagger\end{array}\right)\;. \label{q13}
        \ee
The latter relation together with Eq.~(\ref{q6}) and the identities:
        \be
        ker (D_i^\dagger D_i)=ker(D_i)\;,~~~
        ker (D_i D_i^\dagger)=ker(D_i^\dagger)\;,\label{q100}
        \ee
lead to the desired result \cite{w1}, namely
        \be
        {\rm index}_W=dim(ker\, D_i)-dim(ker\, D_i^\dagger)\;,
        \label{q14}
        \ee
for either of $i=1,2$. In fact, Witten chooses $D_1=D_2$ to
satisfy the compatibility conditions~(\ref{q9})--(\ref{q12}).
If now one identifies ${\cal H}_\pm$ with abstract inner product
(Hilbert) spaces $\Gamma_1$ and $\Gamma_2$, and $D_i:\Gamma_1\to
\Gamma_2$ with two (parameter dependent) Fredholm operators, then
Eq.~(\ref{q14}) implies:
        \be
        {\rm index}_W=\mbox{index}^{\rm Analytic}(D_i)
        \;,
        \label{15}
        \ee
for both $i=1,2$. In particular, one can choose $\Gamma_a$ ($a=1,2$)
to be spaces of smooth sections of a pair
of complex Hermitian vector bundles $E_a$ and $D_i$ a pair of elliptic
differential operators. Then, one has:
        \be
        {\rm index}_W =\mbox{index}^{\rm Atiyah-Singer}(D_i)
        \;,\label{q16}
        \ee
where by the Atiyah-Singer index, we mean the {\em topological index}
introduced by Atiyah and Singer \cite{shanahan}.
Eq.~(\ref{q16}) is proven for twisted Dirac operators and other
classical elliptic operators using the path integral techniques.
The former result together with a result of K-theory lead to a proof of
the general index theorem, \cite{a-g-w,jmp94}.


\section{Parasupersymmetric Topological  Invariant}
In Ref.~\cite{ali}, a  detailed analysis of both the R-S and the B-d
\psqm~ is presented. Here the relevant results are quoted without
proof for brevity.

Consider the R-S parasuperalgebra (\ref{q0.1}), (\ref{q0.2}) written in
terms of the self-adjoint generators (\ref{q4}):
        \bea
        Q_1^3-\{Q_1,Q_2^2\}-Q_2Q_1Q_2=0&&\label{q22}\\
        Q_2^3-\{Q_2,Q_1^2\}-Q_1Q_2Q_1=0&&\label{q23}\\
        \left[Q_1,H\right]=\left[Q_2,H\right]=0 &&\label{q24}\\
        Q_1^3=2Q_1H&&\label{q25}\\
        Q_2^3=2Q_2H&&\label{q26}\;.
        \eea
These relations together with Eq.~(\ref{q4}) which takes the form:
        \be
        H=\frac{1}{4}\left[ (Q_1^2+Q_2^2)^2-3[Q_1,Q_2]^2\right]^{
        \frac{1}{2}}\;,
        \label{q94}
        \ee
and Eqs.~(\ref{q1.5})--(\ref{q1.7}) lead to the following results:
        \begin{itemize}
        \item[1)] The spectrum is non-negative.
        \item[2)] Every parafermionic positive energy state is
        accompanied with a pair of parabosonic states of the
        same energy.\footnote{Note that here one means by the parafermionic
        and parabosonic states the states associated with the
        subspaces ${\cal H}_-$ and ${\cal H}_+$ defined by $\tau$
        through Eq.~(\ref{q3}).} Here one also uses the assumption
        that the involution operator $\tau$ is independent of the
        details of the dynamics, i.e., the Hamiltonian. See \cite{ali} for
        more details.
        \item[3)] In a basis which diagonalizes both $H$ and $Q_1$, one
        has the following representations for $Q_1$, $Q_2$, $\tau$, and
        $H$
        \bea
        \left. Q_1\right|_{{\cal H}_E}&=&\sqrt{2E}\left(
        \begin{array}{ccc}
        1&0&0\\
        0&0&0\\
        0&0&-1\end{array}\right)=\sqrt{2E}J_3^{(1)}\;,\nn\\
        \left. Q_2\right|_{{\cal H}_E}&=&\sqrt{E}\left(
        \begin{array}{ccc}
        0&1&0\\
        1&0&1\\
        0&1&0\end{array}\right)=\sqrt{2E}J_1^{(1)}\;,
        \label{q29}\\
        \left. \tau\right|_{{\cal H}_E}&=&\left(
        \begin{array}{ccc}
        0&0&-1\\
        0&1&0\\
        -1&0&0\end{array}\right)\;,\nn\\
        \left. H\right|_{{\cal H}_E}&=&E\left(
        \begin{array}{ccc}
        1&0&0\\
        0&1&0\\
        0&0&1\end{array}\right)\;,\nn
        \eea
where $J_i^{(1)}$ are the ($j=1$)--representations of the generators of
$SU(2)$.

Switching to a basis which diagonalizes $\tau$ and $H$, one has:
        \bea
        \left. Q_1\right|_{{\cal H}_E}=\sqrt{2E}\left(
        \begin{array}{ccc}
        0&0&0\\
        0&0&1\\
        0&1&0\end{array}\right)&,&
        \left. Q_2\right|_{{\cal H}_E}=\sqrt{2E}\left(
        \begin{array}{ccc}
        0&0&1\\
        0&0&0\\
        1&0&0\end{array}\right)\nn\\
        &&\label{q30}\\
        \left. \tau\right|_{{\cal H}_E}=\left(
        \begin{array}{ccc}
        1&0&0\\
        0&1&0\\
        0&0&-1\end{array}\right)&,&
        \left. H\right|_{{\cal H}_E}=E\left(
        \begin{array}{ccc}
        1&0&0\\
        0&1&0\\
        0&0&1\end{array}\right)\;.
        \nn
        \eea
        \item[4)] In view of Eq.~(\ref{q29}), $Q_1$, $Q_2$, and $H$
        also satisfy the B-D ($p=2$)--superalgebra (\ref{q0.4}), and
        consequently the following simpler set of relations:
        \bea
        Q_1Q_2Q_1&=&Q_2Q_1Q_1\:=\:0\label{q11.0}\\
        \{Q_1,Q_2^2\}&=&Q_1^3\label{q12.0}\\
        \{Q_2,Q_1^2\}&=&Q_2^3\label{q13.0}\\
        2Q_1H&=&Q_1^3\label{q12.1}\\
        2Q_2H&=&Q_2^3\;.\label{q13.1}
        \eea
        \end{itemize}

Items (1) and (2) imply the topological invariance of $\Delta^{(p=2)}$
of Eq.~(\ref{q0.5}). The argument is identical with the one presented for
verifying topological invariance of the Witten index. Namely, under a
continuous deformation of the parasupersymmetric system the
energy levels may move arbitrarily but continuously. In this process
some of the positive energy states may collapse to the zero
level or some zero-energy states may elevate to positive energies.
However, these are only possible if the degeneracy structure is
preserved. This constraint implies invariance of $\Delta^{(p=2)}$
of Eq.~(\ref{q0.5}) under the deformation. The latter is also implicit
in the form of the representation of $\tau$ for positive energy levels
as depicted in Eq.~(\ref{q30}).

Furthermore, item (3) may be employed to obtain an algebraic expression
for the topological invariant. In order to derive such an expression, first
consider the following representation of the Hilbert space:
        \be
        {\cal H}=\left(\begin{array}{c}
        {\cal H}^1_+\\{\cal H}^2_+\\{\cal H}_-\end{array}
        \right)\;,\:\:\:{\rm with}~~~{\cal H}_+=:
        \left(\begin{array}{c}
        {\cal H}^1_+\\{\cal H}^2_+\end{array}\right)\;.
        \label{q32}
        \ee
In view  of the constructions (\ref{q8}) and Eqs.~(\ref{q30}), we propose:
        \be
        Q_1=\left(\begin{array}{ccc}
        0&0&0\\
        0&0&D_1^\dagger\\
        0&D_1&0
        \end{array}\right)\;\;,\;\;
        Q_2=\left(\begin{array}{ccc}
        0&0&D_2^\dagger\\
        0&0&0\\
        D_2&0&0
        \end{array}\right)\;,
        \label{q33}
        \ee
where $D_i:{\cal H}^i_+ \to{\cal H}_-$ ($i=1,2$) are linear operators.
Next we substitute the ansatz (\ref{q33}) in the parasuperalgebra
(\ref{q11.0})--(\ref{q13.1}) and Eq.~(\ref{q94}) for the Hamiltonian.
It turns out that Eqs.~(\ref{q11.0}) are automatically satisfied, whereas
Eqs.~(\ref{q12.0}) and (\ref{q13.0}) lead to the following
compatibility conditions:
        \be
        (D_2D_2^\dagger-D_1D_1^\dagger)D_i=0~~~~
        ~~(i=1,2)\;.
        \label{q34}
        \ee
Imposing Eq.~(\ref{q34}) on (\ref{q94}) leads to considerable simplifications
in the form of the Hamiltonian. One finds:
        \be
        H=\left(
        \begin{array}{ccc}
        D_2^\dagger D_2&0&0\\
        0&D_1^\dagger D_1&0\\
        0&0&D_1D_1^\dagger +D_2D_2^\dagger
        \end{array}
        \right) \;.\label{q36}
        \ee
In view of (\ref{q34}) and (\ref{q36}), Eqs~(\ref{q12.1}) and (\ref{q13.1})
also are satisfied as identities.

Finally, using Eq.~(\ref{q36}), one can easily derive the desired
expression for $\ddd$:
        \be
        \ddd=dim(ker~D_1)+dim(ker~D_2)-2dim(ker~D_1^\dagger
        \cap ker~D_2^\dagger)\;.
        \label{q39}
        \ee
Here we have employed the following identifications:
        \bea
        n_0^{\pi B}&=&dim(ker~D_1^\dagger D_1\oplus ker~D_2^\dagger D_2)
        \nn\\
                   &=&dim(ker~D_1)+dim(ker~D_2)\label{q37}\\
        n_0^{\pi F}&=&dim (ker~[D_1\,D_1^\dagger+D_2\,D_2^\dagger])\nn\\
        &=&dim(ker~D_1^\dagger \cap ker~D_2^\dagger)\;.
        \label{q38}
        \eea
In Eqs.~(\ref{q37}) and (\ref{q38}) use is made of relations~(\ref{q100}).

It turns out that conditions (\ref{q34}) may be used to simplify the
expression for $\ddd$. To see this let us define $A_i:=D_iD_i^\dagger$,
(i=1,2). Then multiplying Eqs.~(\ref{q34}) by $D_i^\dagger$ from
the right and writing the resulting equations in terms of $A_i$, one
has:
        \be
        (A_1-A_2)A_1=0~,~~~~(A_1-A_2)A_2=0\;.
        \label{q120}
        \ee
In view of the fact that $A_i$ are self-adjoint, positive definite operators,
Eqs.~(\ref{q120}) imply $ker~A_1=ker~A_2$. This together with the
identities~(\ref{q100}), leads to $ker~D_1^\dagger=ker~D_2^\dagger$.
Thus, we have:
        \be
        \ddd={\rm index}^{\rm analytic}(D_1)+
        {\rm index}^{\rm analytic}(D_2)\;.
        \label{q121}
        \ee
Eq.~(\ref{q121}) provides the desired mathematical interpretation for the
parasupersymmetric topological invariant considered in this letter.

\section{Conclusion}
We conclude this letter by remarking that the introduction of $\ddd$
directly depended on the choice of the Hamiltonian, i.e.,
Eq.~(\ref{q94}). One may try to investigate other possible forms
of the Hamiltonian which are compatible with the defining parasuperalgebras
of ($p=2$)--PSQM, and even attempt to classify the corresponding systems
and their topological invariants. Another possible direction of
further research is to investigate the topological aspects of
PSQM of orders: $p>2$.

\section*{Acknowledgements}
The auther would like to thank Drs.~V.~Karimipour, S.~Rouhani, and
A.~Rezaii for invaluable comments and discussions.

\end{document}